\documentstyle[amsfonts,amssymb,12pt,epsf]{article}

\makeatletter
\@addtoreset{equation}{section}

\newcommand{\nn}{\nonumber}
\newcommand{\vs}[1]{\vspace*{#1}}
\newcommand{\hs}[1]{\hspace*{#1}}
\newcommand{\tr}{\mathop{\rm Tr}}

\newcommand{\Half}{\frac12}
\newcommand{\unit}{\hbox to 3.8pt{\hskip1.3pt \vrule height 7.4pt
    width .4pt \hskip.7pt \vrule height 7.85pt width .4pt \kern-2.4pt 
    \hrulefill \kern-3pt \raise 3.7pt\hbox{\char'40}}}

\def\href#1#2{#2}

\def\a{\alpha'}
\def\F{{\bf F}}
\def\B{{\bf B}}
\def\Y{{\bf Y}}
\def\A{{\bf A}}
\def\dv{{\bf v}^\dagger}
\def\v{{\bf v}}
\newcommand{\R}{{\Bbb R}}

\begin{document}


\begin{titlepage}


\title{
\vs{-10mm}
\hfill\parbox{4cm}{
{\normalsize\tt hep-th/0101145}
}
\\[10pt]
D-brane Solutions in Non-Commutative Gauge Theory \\
on Fuzzy Sphere
}
\author{
Koji {\sc Hashimoto}\thanks{{\tt 
koji@itp.ucsb.edu}}$\hspace{2mm}{}^a$
and 
Kirill {\sc Krasnov}\thanks{{\tt 
krasnov@cosmic.physics.ucsb.edu}}$\hspace{2mm}{}^b$
\\[10pt]
${}^a${\it Institute for Theoretical Physics,}\\
{\it University of California, Santa Barbara, CA 93106}\\
${}^b${\it Department of Physics, }\\
{\it University of California, Santa Barbara, CA 93106}\\
}
\date{\normalsize January, 2001}
\maketitle
\thispagestyle{empty}

\begin{abstract}
\normalsize\noindent Non-commutative gauge theory on fuzzy sphere
was obtained by Alekseev {\it et al.} as describing the low energy 
dynamics of a spherical D2-brane in $S^3$ with the background
$b$-field. We identify a subset of solutions of this theory which are
analogs of ``unstable'' solitons on a non-commutative  flat D2-brane
found by Gopakumar {\it et al}. Analogously to the flat case, these
solutions have the interpretation as describing D0-branes ``not yet    
dissolved'' by the D2-brane. We confirm this interpretation by showing
the precise agreement of the binding energy computed in the
non-commutative and ordinary Born-Infeld descriptions. We then study
stability of the solution describing a single D0-brane off a
D2-brane. Similarly to the flat case, we find an instability when the
D0-brane is located close to the D2-brane. We furthermore obtain the
complete mass spectrum of 0-2 fluctuations, which thus gives a
prediction for the low energy spectrum of the 0-2 CFT in $S^3$. We
also discuss in detail how the instability to a formation of 
the fuzzy sphere modifies the usual Higgs mechanism for small 
separation between the branes.
\end{abstract}

\end{titlepage}


\section{Introduction}

In recent years, much of the progress of string theory was based on a
study of statics and dynamics of D-branes in various situations. A
particularly interesting recent development is the introduction
of non-commutativity into the D-brane worldvolume theory
\cite{Connes,doug,SW}. The non-commutativity 
makes transparent the relation between D-branes of different
worldvolume dimensions through the construction of various solitons
\cite{HK,Nikita,Nikita-Dynamics,Strom,Nikita-Solitons},  
which can be thought of as a manifestation of the ``brane democracy'' 
\cite{Townsend}. 

So far, most of the attention was given to the case of constant
non-commutativity, which is what one has for the flat D-branes in 
a non-varying NS-NS 2-form field. Potentially interesting is the more
general case of a non-vanishing field strength. Such backgrounds are
necessarily curved, which makes the analysis more difficult. The
potential payoff here is that one may be able to observe aspects of
brane dynamics which are absent in the flat case.

One of the simplest backgrounds of this type is $S^3\times M_7$, where
$M_7$ is some 7-dimensional manifold. This background can be realized in 
string theory as the near-horizon geometry of a stack of NS5-branes.
There exists an exact description of D2-branes on $S^3$ in terms of
${\rm SU}(2)$ WZW model. One finds, see
\cite{Alekseev-Old-1,Alekseev-Old-2,Bachas,Paw},  
that supersymmetric D2-branes wrap certain conjugacy
classes in ${\rm SU}(2)$. These are certain integral spheres in $S^3$.
For radii of the spherical D2-brane much smaller than the radius of
curvature of $S^3$, and in the Seiberg-Witten (SW) limit $\a\to 0$,
the geometry of D2-brane worldvolume becomes \cite{Alekseev-Old-2}
that of a fuzzy sphere \cite{Madore}. As was shown in \cite{Alekseev},
the low-energy dynamics in this case is described by a certain
non-commutative gauge  theory on fuzzy $S^2$. The action of this
theory contains, apart from the usual Yang-Mills (YM) term, the
Chern-Simons (CS) term. This is one of the examples in which YM-CS
theory appears in string theory context \cite{Kitao, Imamura}.

A large set of solutions of this theory was obtained in
\cite{Alekseev}. The solutions were interpreted as describing stacks
of D2-branes of various radii. As we show in this paper, there is a
particular interesting and simple subset of solutions of
\cite{Alekseev} that describe a single D2-brane together with a number
of D0-branes. These solutions are exact analogs (in 2+1 dimensions) of
the non-commutative monopole solution found in \cite{Nikita} and of
the ``unstable'' solitons that were found in \cite{Strom} for the case
of a flat D2-brane. We describe this set of solutions, and confirm
their D0-brane interpretation by comparing the energy obtained in the
non-commutative gauge theory to that found in the ordinary Born-Infeld
(BI) description. 

The non-commutative solution describing co-centric D2-brane 
shells was argued to be stable in \cite{Alekseev}. In the 
present paper we extend the analysis of stability to the case of 
non-cocentric shells, considering the subset of solutions that have
the D0-brane interpretation. We find that, when a D0-brane is located
too close to the shell of the D2-brane, the system becomes unstable
to forming a larger D2-brane. Furthermore, we
completely diagonalize the set of 0-2 fluctuation modes. We find a
discrete, finite set of massive modes. This gives a prediction for 
the spectrum of light modes of the 0-2 worldsheet CFT on $S^3$ in the 
SW limit.

Our results on stability of the configuration consisting of a single
D2-brane together with a D0-brane have interesting applications
for the phenomenon of gauge symmetry breaking. Consider D$p$-branes
put on $S^3\times M_7$ in such a way that the directions along the
branes are all in $M_7$. In the usual flat case D-brane Higgs
mechanism the gauge symmetry on the D$p$-brane worldvolume is broken
by separating the branes, and the masses of gauge bosons are
proportional to the distance between them \cite{Witten}. In our case,
the usual mechanism is valid for large enough separation of the branes
in $S^3$. Thus, for two D$p$-branes that are separated in $S^3$ by a
large enough distance the worldvolume gauge symmetry is broken from
${\rm U}(2)$ to ${\rm U}(1)$, as usual. However, when the branes are
located too close, this usual mechanism is no longer valid. Indeed, as
our analysis shows, two D0-branes in $S^3$ located too close will form
a fuzzy sphere (of smallest non-trivial radius). This breaks the ${\rm
  U}(2)$ symmetry completely. Thus, as the separation between the
branes decreases one gets a complete symmetry breaking, instead of the
expected restoration of symmetry, as in the flat case. Therefore the
phenomenon of polarization of branes \cite{Emparan,Myers, Park} leads
to a complete breaking of the worldvolume gauge symmetry and in
particular prevents the symmetry restoration. 

The organization of the paper is as follows. In the next section,
for readers convenience, we review some known material. We first give,
following \cite{Bachas} a brief account of the ordinary BI
(Born-Infeld) description of spherical D2-brane in $S^3$ in the
background $b$-field. The second part of this section reviews some
facts about the non-commutative gauge theory of \cite{Alekseev}. Here
we give the action and describe what is known about the solutions. In
section \ref{sec:norm} we fix the coefficient in front of the action
of \cite{Alekseev}, which is left unspecified in that paper. We shall
need the precise normalization of the action to compare the energy
calculated in the commutative and non-commutative descriptions. Section
\ref{sec:sol} describes in detail the set of solutions in question and
gives their interpretation in terms of D0-branes. In section
\ref{sec:stab} we study fluctuations about these solutions, analyze
stability and obtain the mass spectrum of fluctuations. In section
\ref{sec:higgs} we discuss our results in view of the phenomenon of
symmetry breaking. We conclude with a discussion.

While writing this paper we received the paper \cite{Japan},
which also considers the non-commutative gauge theory on fuzzy sphere,
and studies fluctuations around certain solutions (one describing a
single D2-brane and one describing two D0-branes). Since no comments
are made in that paper as to the stability of the later system, the
overlap with the present paper is rather marginal.


\section{Review: Ordinary BI Description, Non-Commutative Gauge 
Theory on Fuzzy Sphere and Solutions}
\label{sec:review}

A spherical D-brane tends to shrink to minimize its energy. However, 
in the presence of some background fields that couple to the brane,
a stabilization mechanism is known to work
\cite{Emparan,Myers,Bachas,Paw,Park}. The simplest case is that of a
D2-brane in the background of either a non-trivial R-R 3-form field,
or NS-NS 2-form field with a non-vanishing field strength. The later
situation is realized in the background $S^3\times M_7$, which appears
as the near-horizon geometry of a stack of NS5-branes. Since there
exists the NS-NS $b$-field flux on $S^3$, the D2-branes on it are
possibly stabilized with a definite extent. D2-branes on $S^3$ were
studied in the ordinary BI description in
\cite{Bachas,Paw}, and using a non-commutative worldvolume theory in
\cite{Alekseev-Old-2,Alekseev}. Here we review some of the facts that
will be needed in the following. 


\subsection{Ordinary BI description}

In this subsection, we shall review Ref.\ \cite{Bachas} briefly, which
treats D2-branes in $S^3$ using BI theory (see also \cite{Paw}). The
metric on $S^3$ is given by
\begin{equation}
ds^2 = k\a \left(
d\psi^2 + \sin^2\psi(d\theta^2+\sin^2\theta d\varphi^2)
\right).
\label{metric}
\end{equation}
Here $k$ is the number of NS5-branes that were used to obtain the
background. Then $k\a$ is the radius of $S^3$. The NS-NS 2-form field
strength is proportional to the volume form of $S^3$ and is given by
\begin{equation}
H = 2 k\a \sin^2\psi \sin\theta d\psi\wedge d\theta\wedge d\varphi.
\end{equation}
The corresponding NS-NS 2-form potential $b: db = H$ is given by
\begin{equation}
2\pi\a b = k\a \left( \psi - {\sin2\psi\over 2} \right) \sin\theta 
d\theta\wedge d\varphi.
\label{B}
\end{equation}
The D2-branes in this background are stabilized by the flux of the
${\rm U}(1)$ gauge field. This flux can take integral values
corresponding to different ${\rm U}(1)$ bundles one can have over
$S^2$. Thus, 
\begin{equation}
F = - {n\over 2} \sin\theta d\theta\wedge d\varphi,
\label{F}
\end{equation}
so that
\begin{equation}
\int_{S^2} F = - 2\pi n.
\end{equation}
The energy of a spherical D2-brane with $n$ units of flux on it 
located at $\psi={\rm const}$ is then given by the BI action. 
Substituting (\ref{metric}), (\ref{B}) and (\ref{F}) into the 
D2-brane action, one gets for the energy 
\begin{equation}
E_n(\psi) = k\a T_{(2)} \int d\theta d\varphi \sin\theta
\sqrt{\sin^2\psi + \left(\psi - {\sin2\psi\over 2} - {\pi n\over k}
\right)^2}.
\end{equation}
Here $T_{(2)}$ is the D2-brane tension. The energy is minimized by 
\begin{equation}
\psi_n = {\pi n\over k}.
\label{psin}
\end{equation}
The corresponding energy is
\begin{equation}
E_n \equiv E_n(\psi_n) = 4\pi k\a T_{(2)} \sin{\pi n\over k}.
\label{energyen}
\end{equation}
Thus, the stable configuration, corresponding to the minimum of the 
energy, is a spherical D2-brane, whose radius depends on the amount of
the ${\rm U}(1)$ flux on it. We will use this formula for the energy
in section \ref{sec:sol}.


\subsection{Non-commutative description}

Here we review Ref.\ \cite{Alekseev} in the amount we need for the
following. According to this paper, the low energy effective theory on
a stack of $N$ D2-branes of size $n$ is given, in the $n/k \ll 1$
limit, by a non-commutative ${\rm U}(N)$ gauge theory  on a fuzzy
sphere of ``radius'' $n$. The action, as proposed by  \cite{Alekseev},
is given by the sum of the YM and the CS terms. The YM term reads 
\begin{equation}
\label{action-YM}
S_{\rm YM} = {1\over 4} (2\pi\a)^2 {1\over {\rm Dim}} {\rm Tr} \F_{ab}
\F^{ab}. 
\end{equation}
Here one raises and lowers indices using the metric ${\bf g}^{ab} =
(2/k) \delta^{ab}$. We have restored the factor of $(2\pi\a)^2$ put to
unity in \cite{Alekseev}, and explicitly specified that one uses the 
normalized trace. The curvature $\F^{ab}$ of a non-commutative 
${\rm U}(N)$ connection ${\bf A}^a$ is most easily expressed in terms  
of the ``covariant coordinates'' \cite{Madore-Gauge} $\B^a$
\begin{equation}
{\bf B}^a = {\bf Y}^a + {\bf A}^a.
\end{equation}
Here ${\bf Y}^a = Y^a/\sqrt{2\a}$, and $Y^a$ are the fuzzy sphere 
coordinates satisfying the usual ${\rm SU}(2)$ commutation relations
\begin{eqnarray}
[Y^a,Y^b]=i\epsilon^{abc} Y^c.
\label{yabc}
\end{eqnarray}
Here $\epsilon^{abc}$ is the Levi-Civita symbol, and the
contraction over the repeated index is $c$ assumed. 
Using the covariant coordinate ${\bf B}^a$, the field strength
can be written as
\begin{equation}
{\bf F}^{ab} = i [\B^a, \B^b] + {\bf f}^{ab}_{\quad\! c} \B^c,
\end{equation}
where ${\bf f}^{ab}_{\quad\! c} = (2/k) \epsilon^{ab}_{\quad\! c} 
/\sqrt{2\a}$. The Chern-Simons term is best written in terms of the
covariant coordinate $\B^a$ and reads
\begin{equation}
S_{\rm CS}= - {i\over 2} (2\pi\a)^2 {1\over {\rm Dim}} {\rm Tr}\left(
{1\over 3} {\bf f}^{abc} \B_a \B_b \B_c - {i\over \a k} \B_a \B^a
+ {i\over 3\a k} \Y_a \Y^a \right).
\end{equation}
The full action, written in terms of the covariant coordinate $\B^a$, 
then takes the form
\begin{equation}
\label{action-full}
(2\pi\a)^2 {1\over {\rm Dim}} {\rm Tr} \left(
- {1\over 4} [\B_a,\B_b] [\B^a,\B^b] 
+ {i\over 3} {\bf f}^{abc} \B_a [\B_b,\B_c]
+ {1\over 6\a k} \Y_a \Y^a \right).
\end{equation}
Note that the last term is exactly such that the action takes zero
value when $\B^a$ is an irreducible representation $\B^a=\Y^a$, that
is on the vanishing connection.

Having described the action for the non-commutative gauge theory in
question, let us review what is known about its solutions. The
equations of motion that one derives from (\ref{action-full}) read
\begin{equation}
\label{equations}
[\B^a,\F_{ab}] = 0.
\end{equation}
If one uses this gauge theory to describe a stack of $N$ D2-branes
wrapping a fuzzy sphere of radius $n$, the ``fields'' entering the
action and equations of motion are just matrices in 
${\rm Mat}_N\otimes{\rm Mat}_n$, and $Y^a = {\bf 1}_N\otimes Y^a_n$,
where $Y^a_n$ are generators in the $n$-dimensional irreducible
representation. Ref.\ \cite{Alekseev} describes a large set of
solutions to (\ref{equations}). These are configurations for which the
gauge field $\A$ is constant $\A_a={\bf S}_a$, that is, commutes with
the fuzzy sphere coordinates: $[\Y_a,{\bf S}_b]=0$. There are two
types of such solutions: (i) ${\bf S}_a$ is any set of commuting
matrices from ${\rm Mat}_N\otimes{\bf 1}_n$. These matrices can then
be diagonalized; eigenvalues have the obvious meaning of the
coordinates of the centers of $N$ fuzzy spheres in $S^3$. (ii) ${\bf
  S}_a\in {\rm Mat}_N\otimes {\bf 1}_n$ is any, not necessarily
irreducible, representation of ${\rm SU}(2)$, so that it satisfies 
the usual commutation relation (\ref{yabc}). Note that this type of
solutions corresponds to flat connections $\F_{ab}=0$. According to
\cite{Alekseev}, such a solution corresponds to a fixed point of an RG
flow that takes one from the original configuration of $N$ branes of
radius $n$ centered around the origin to a new configuration
consisting of branes of various radii, also centered around the
origin. These branes can all have different radii, or some of them can
be stacked, or the new configuration can be just a single brane. For
the precise recipe for determining the final configuration see
\cite{Alekseev}, formula (5.4). We shall give a somewhat different,
although equivalent, description of a subset of the above solutions in
section \ref{sec:sol}. As a final remark of this section, let us note
that solutions for which $\B^a$ is a representation (not necessarily
irreducible), that is satisfy the commutation relation (\ref{yabc})
are argued to be stable in \cite{Alekseev}. These solutions describe
flat connections. We shall analyze the stability of certain
interesting non-flat solutions in section \ref{sec:stab}.


\section{Normalization of the Action}
\label{sec:norm}

Before we describe a subset of solutions that has a simple
interpretation in terms of D0-branes, let us discuss the normalization
of the action (\ref{action-full}). It is clear that some factor of the
brane tension must be included in the front. To fix it, we shall
compare the YM part (\ref{action-YM}) of the action with the
Yang-Mills action that can be obtained from the non-commutative BI
action. 

Let us, however, first rescale the fields for future convenience. We
introduce  new gauge field $A = \sqrt{2\a} {\bf A}$ and the
corresponding covariant  coordinate field $B$ and curvature $F$. Let
us also lower the indices of one of the $F$'s in the action using the
${\bf g}$ metric. We get 
\begin{equation}
\label{YM-al}
S_{\rm YM} = {(2\pi\a)^2\over (k\a)^2} {1\over 4} {1\over {\rm Dim}}
{\rm Tr} F_{ab} F_{ab},
\end{equation}
with the contraction over repeated indices assumed.

To fix the prefactor in front of this action, which, as we shall see, 
is proportional to a certain multiple of the $T_{(2)}$ (or $T_{(0)}$) 
brane tension, we would like to compare the above action with the
one that can be obtained from the non-commutative BI action. The
standard calculation gives
\begin{equation}
S_{\rm YM} = \hat{T}_{(2)} (2\pi\a)^2 \int\! d^2x \sqrt{P(G)} 
\;\frac14
{\rm Tr} F_{ab} F^{ab}.
\end{equation}
Here $P(G)$ is the determinant of the pullback of the open string 
metric $G_{ab}$ on $S^2$. We have also introduced the
``non-commutative'' D2-brane tension
\begin{equation}
\hat{T}_{(2)} = {1\over G_{\rm s} (2\pi)^2 (\a)^{3/2}},
\end{equation}
where $G_s$ is the open string coupling constant. Let us relate it to 
the closed string coupling constant $g_s$. A simple comparison of the  
ordinary and non-commutative BI actions gives \cite{SW}
\begin{equation}
G_{\rm s} = g_{\rm s} 
\left ( {{\rm det} P(G)\over {\rm det} P(g+ 2\pi\a(b+F))} 
\right)^{1/2},
\end{equation}
where $P$ is the projection onto $S^2$, and $G$ is the open string
metric.  To calculate the right hand side we need to know $G$. This
can be done using the formula for $G$ in terms of $g, b+F$. As is
shown in \cite{SW}, this relation is as follows
\begin{equation}
G^{ab} = \left( {1\over g+2\pi\a(b+F)} \right)^{ab}_{\rm S},
\end{equation}
where the subscript S denotes the symmetric part. Using the
explicit expression for the closed string metric and the 2-form
fields from section \ref{sec:review}, one finds
\begin{equation}
\left( {1\over g+2\pi\a(b+F)} \right)_{\psi=\psi_n} = {1\over k\a} 
\left(
\begin{array}{ccc}
1 & 0 & 0 \\
0 & 1 & {\cos\psi_n\over \sin\psi_n} {1\over \sin\theta} \\
0 & - {\cos\psi_n\over \sin\psi_n} {1\over \sin\theta} & 
{1\over \sin^2\theta}
\end{array} \right).
\end{equation}
Thus, the restriction $P(G)$ of the open string metric on $S^2$
is $k\a$ times the canonical metric on $S^2$. 
Therefore, the quantities necessary for the computation of the action
are obtained as
\begin{equation}
\sqrt{{\rm det} P(G)} = k\a \sin\theta,
\end{equation}
and
\begin{equation}
G_s = {g_s\over \sin{\pi n\over k}},
\end{equation}
where we have substituted the expression for $b+F$ at
$\psi=\psi_n$. We see that the relation between open and closed string
coupling constants depends on the size $n$ of the fuzzy sphere. 
This is clearly due to the fact that the NS-NS 2-form field is 
varying in $S^3$. As is discussed in \cite{Alekseev}, the fuzzy
sphere description is only valid in the limit $n/k \ll 1$. In this
regime the relation between coupling constants becomes
\begin{equation}\label{relation}
G_s = {g_s k\over \pi n}.
\end{equation}

Having found the relation between the closed and open string coupling
constants, it is easy to fix the coefficient in front of the action.  
To go from the integral over the $S^2$ to the trace of the fuzzy
sphere we have to use the prescription
\begin{equation}
{1\over 4\pi k\a} \int\! \sqrt{P(G)}\; 
f \to {1\over {\rm Dim}} {\rm Tr}
\hat{f}. 
\end{equation}
Here $f$ is a function on $S^2$ and $\hat{f}$ is the corresponding
matrix representation on the fuzzy sphere, and $\rm Dim$ is the
dimension of the irreducible representation of ${\rm SU}(2)$ that is
used to construct the fuzzy sphere. The map from the integral to the
trace is normalized so that both sides coincide when $f=1$. Using this 
prescription we can write
\begin{equation}
S_{\rm YM} = 4\pi k\a \hat{T}_{(2)} (2\pi\a)^2 {1\over {\rm Dim}}
{\rm Tr} {1\over 4} F_{ab} F^{ab}.
\end{equation}
Here the indices are raised and lowered using the open string metric
$G$. Using (\ref{relation}), we can now rewrite the 
action in terms of the closed string
coupling constant and the flat metric. We get
\begin{equation}
S_{\rm YM} = n T_{(0)} (2\pi\a)^2 {1\over (k\a)^2} {1\over {\rm Dim}}
{\rm Tr} {1\over 4} F_{ab} F_{ab}.
\end{equation}
This is to be compared with (\ref{YM-al}). We see that what we have
obtained is essentially the action (\ref{YM-al}), apart from the
prefactor of $n T_{(0)}$. In fact, this prefactor could have been
guessed even without the above derivation. Indeed, the $n$ cancels
with the factor of $1/{\rm Dim}$ normalizing the trace, and what one
is left with is just the action for matrix model of D0-branes, with 
the correct factor of $T_{(0)}$ in front.

Thus, adding the CS term with the same normalization coefficient, we
obtain the full expression for the action as
\begin{eqnarray}
  S[B] = 
n T_{(0)} \left({2\pi\over k}\right)^2 {1\over {\rm Dim}}\tr
  \left(
    -\frac14 [{B}_a, {B}_b]^2 
    + \frac{i}{3}\epsilon_{abc}{B}_a [{B}_b, {B}_c]
    +\frac16 {Y}_a^2
  \right),
\label{redefac}
\end{eqnarray}
Here ${Y}_a$ satisfy the $SU(2)$ algebra (\ref{yabc}).


\section{Solutions and their D-brane Interpretation}
\label{sec:sol}

We are now ready to describe the subset of solutions that have
an interpretation in terms of D0-branes.


\subsection{Subset of solutions}

{}From now on, we will use the action (\ref{redefac}) written in terms 
of rescaled covariant coordinates $B_a$ and with all indices
contracted with the help of the usual flat metric in $\R^3$. The
equations of motion that one derives from this action read
\begin{eqnarray}
\label{eq-motion}
  [{B}_b, F_{ab}]=0,
\end{eqnarray}
where the field strength is 
\begin{eqnarray}
F_{ab}= i[{B}_a, {B}_b]+\epsilon_{abc}{B}_c.
\label{defF}
\end{eqnarray}

A large set of solutions was described in \cite{Alekseev} and reviewed
in section \ref{sec:review} above. As we said in the introduction,
in this paper we would like to consider a particular subset of
solutions, which will later receive the interpretation as describing a
single spherical D2-brane together with a number of D0-branes. 
Restriction to a single D2-brane is for simplicity only and
can be lifted in a straightforward way. The set of solutions of
interest can be constructed by first constructing the one describing
branes centered at the origin and then modifying it to shift the
branes. A configuration with all branes located at the origin
preserves the ${\rm SU}(2)$ rotational symmetry and, thus, must have
the vanishing field strength
\begin{eqnarray}
  F_{ab}=0.
\end{eqnarray}
The definition (\ref{defF}) of the field strength shows that this is
equivalent to the requirement that $B_a$ are generators of some (not
necessarily irreducible) representation of ${\rm SU}(2)$. The case of
$B$ generating an irreducible representation clearly corresponds to a
single spherical D2-brane. The case of $B$ splitting into several
irreducible components corresponds to several spherical D2-branes all
centered at the origin. Among these irreducible components one can
have the trivial representation. It describes D2-branes of zero size,
or, equivalently, D0-branes, as we shall demonstrate below. 

Hence, a solution describing a single D2-brane plus a number of
D0-branes, all centered at the origin is simply
\begin{equation}
B^a = \left(
\begin{array}{cccc}
0 & \ldots & 0 & 0\\
\vdots & \ddots & \vdots & 0 \\
0 & \ldots & 0 & 0 \\
0 & 0 & 0 & Y_n^a 
\end{array} \right).
\end{equation}
Here $Y_n^a$ are generators in the irreducible representation of
dimension $(n+1)$. To get a more general solution with arbitrary brane 
locations we need to shift their positions. We can shift both the
D2-brane and D0-branes, but this is not necessary since we are only
interested in relative configuration of the branes. Thus, we will
always keep the D2-brane centered at the origin. Then the shifting of
D0-branes amounts to replacing zero's on the diagonal with numbers
that receive the interpretation of coordinates of D0-branes in $S^3$:
\begin{equation}
\label{sol-B}
B^a = \left(
\begin{array}{cccc}
c_1^a & \ldots & 0 & 0\\
\vdots & \ddots & \vdots & 0 \\
0 & \ldots & c_l^a & 0 \\
0 & 0 & 0 & Y_n^a 
\end{array} \right).
\end{equation}
The number of D0-branes described is equal to $l$. It is easy to see
that the field strength (\ref{defF}) on (\ref{sol-B}) is non-zero
and is equal to
\begin{equation}
\label{sol-F}
F_{ab} = \epsilon_{abc} \left(
\begin{array}{cccc}
c_1^c & \ldots & 0 & 0\\
\vdots & \ddots & \vdots & 0 \\
0 & \ldots & c_l^c & 0 \\
0 & 0 & 0 & 0 
\end{array} \right).
\end{equation}
This clearly satisfies the field equations, for $F_{ab}$ commutes with
$B^a$. Thus, (\ref{sol-B}) is indeed a solution.

Before we proceed with the discussion as to the properties of this
solution, several comments are in order. First of all, let us note
that, unlike the discussion in \cite{Alekseev}, we choose to describe
the solution directly in terms of the covariant coordinates $B^a$,
not in terms of the gauge field $A^a$. This is more convenient, for
the components of the covariant coordinates $B^a$ receive the direct 
interpretation of coordinates in $S^3$, in the case when $B$ can
be (at least partially) diagonalized. In addition, this description in 
terms of $B^a$ is rather natural, because it makes transparent the
relation to T-dualities in Matrix theory \cite{Taylor2, Kawano}. 
Second, the set of solutions that is 
obtainable from D-branes centered at the
origin by shifts by commuting matrices is, probably, not the most
general set. This way one can only obtain configurations solving
(\ref{eq-motion}) rather trivially, in the sense that $F$ commutes with
$B$. It would be interesting to find examples of solutions, if any,
for which this is not the case.

The solution (\ref{sol-B}) is the direct analog of the ``unstable''
solitons found in the context of flat D2-branes, see
\cite{Strom,Nikita-Solitons}. As is explained in
\cite{Nikita-Solitons}, in non-commutative gauge theory on flat
D2-brane, the requirement of finiteness of energy only allows
configurations in which the covariant derivative $D$ (and its
conjugate $\bar{D}$) generates a reducible representation of the
Heisenberg algebra. The covariant derivative $D$, or, in the
terminology of \cite{Madore-Gauge} covariant coordinate, is an
infinite rank matrix. The requirement of finite energy fixes it to be
of a block diagonal form, in which the first block is diagonal, with
numbers on the diagonal having the interpretation of positions of
solitons, and the second block  filling the usual infinite dimensional
representation of the Heisenberg algebra. Our solution (\ref{sol-B})
falls into the same general scheme: it forms a reducible
representation of the symmetry algebra in question, splitting into a
single irreducible and several trivial components. Like in the flat 
case, the irreducible component describes the D2-brane itself,
and trivial representations correspond to D0-branes. The main 
difference with the flat case is that the D2-brane is
represented by a finite dimensional matrix. Note that one can 
add other non-trivial irreducible representations. This
means that we  can have small spherical D2-branes as ``solitons'' of
the large mother D2-brane, although we do not consider this case in
the present paper. Note that the distinction of which D2-brane is a
``soliton'' and which provides a ``background'' becomes relative. Any
D2-brane can be viewed as the ``mother'' on whose worldvolume the
non-commutative gauge theory lives. This serves as a good illustration
of the ``brane democracy'' \cite{Townsend} in string theory. 
  
Therefore, our solution is the analog of the one found in
\cite{Strom,Nikita-Solitons}, and thus should also describe
D0-branes. We shall directly confirm this interpretation by comparing
the energy found in the ordinary BI and non-commutative
descriptions. However, before we proceed with this comparison of
energies, let us make one more remark. As we have seen, the
configuration corresponding to a single D2-brane has vanishing 
field strength $F_{ab}$ (this also follows from its spherical
symmetry). This is a little puzzling, for the integral of this two
form over the D2-brane has the interpretation of the number of
D0-branes. Indeed, the intuition based on the Myers effect
\cite{Myers} is that the  fuzzy D2-brane worldvolume is ``made out''
of a number of D0-branes. However, by simply integrating the
non-commutative field strength over the D2-brane, which in
non-commutative description is replaced by the operation of taking the
trace, we get zero. The resolution of this puzzle is that in
non-commutative description the number of D0-branes is no longer given
by the integral (trace) of $F$. This interpretation comes in the
commutative case from the CS term  
\begin{equation}
T_p \int\! d^{p+1}x\; \left( \sum C \right) \wedge e^{2\pi\a(b+F)}.
\end{equation}
As it has been discussed in the literature, see, e.g., \cite{CS},
in the non-commutative description the quantity $2\pi\a(b+F)$ in the 
exponential must be replaced by its Seiberg-Witten map \cite{SW} 
image, which is the matrix that was denoted by $Q^{-1}$ in
\cite{CS}. One also gets a factor of ${\rm Pf}Q$ under the trace,
see \cite{CS}. As one can easily check, this gives that the
number of D0-branes is given in non-commutative description 
by the trace of the identity operator, or by the dimension of the 
Hilbert space. This is of course consistent with what we expect
from the Myers effect interpretation of the fuzzy D2-brane, for 
the D2-brane made out of $n$ D0-branes is described by 
matrices of rank $n$. In the flat case, the non-commutative D2-brane
can be thought of as made of an infinite number of D0-branes, and the
operator ${\rm Tr}F$ measures only the number of not yet
dissolved D0-branes, not the total number in the system. It is
instructive to compare the discussion of this paragraph to  
the discussion on the quantization of D0-brane charge appeared
recently in the literature, see, e.g., \cite{Bachas,Taylor,Marolf}.


\subsection{Comparison of the energy}

In order to confirm that the solution (\ref{sol-B}) 
actually corresponds to a set of the D0-branes and a spherical
D2-brane we compare the energy of this solution to the energy that can
be obtained in the ordinary BI description of section
\ref{sec:review}. First, let us obtain the prediction of the ordinary
BI theory.  

We would like to understand the $n$ dependence of the energy 
of the spherical D2-brane (\ref{energyen}) in the region $n/k \ll
1$. Physically, this corresponds to D2-branes of radius much smaller
than the radius of curvature of $S^3$. It is for such branes that the
our description in terms of the gauge theory on a fuzzy
sphere is applicable. We have
\begin{equation}
E_n = 4\pi k\a T_{(2)} \left[ \left({\pi n\over k}\right) -
{1\over 3!} \left({\pi n\over k}\right)^3 \right] 
+ {\cal O}\left((n/k)^5\right) =
T_{(0)} n \left[ 1 -
{1\over 3!} \left({\pi n\over k}\right)^2 \right] 
+ {\cal O}\left((n/k)^5\right) ,
\end{equation}
where we have used the fact that $4\pi^2 \a T_{(2)} = T_{(0)}$.
The first term has an obvious interpretation as the energy of $n$
D0-branes. The second term can be interpreted as the binding energy
which 
is released when the $n$ D0-branes form a single D2-brane. Thus, in
the limit $n/k \ll 1$ we can write 
\begin{equation}
E_{n {\rm D0}} - E_n = T_{(0)} n
{1\over 3!} \left({\pi n\over k}\right)^2.
\label{diff1}
\end{equation}
As we shall see in a moment, this energy difference is reproduced
by the non-commutative description. 

Another quantity that we need is the binding energy of a single
D0-brane and a D2-brane. This can be derived analogously to what was
done in Ref.\ \cite{Strom}. One expands the BI action in the limit of
large $b+F$ field. The binding energy is then read from the variation
of the total energy when $b+F$ changes so that
\begin{equation}
\int \delta(b+F) = 2\pi,
\end{equation}
which corresponds to the emission of a single D0-brane. Completely
analogously 
to the flat case, we get
\begin{equation}
E_{\rm bind} = {T_{(0)}\over 2} \left(\sqrt{g}\over 2\pi\a (b+F)\right).
\end{equation}
Here $\sqrt{g}$ is the square root of the determinant of the induced
(closed string) metric on the D2-brane, and $b+F$ is the
$(\theta\varphi)$-component of the $b+F$ 
two-form. Substituting the values of these quantities at $\psi_n$ we
get
\begin{equation}
E_{\rm bind} = {T_{(0)}\over 2} \tan^2{\pi n\over k}\approx
{T_{(0)}\over 2} \left({\pi n\over k} \right)^2,
\label{diff2}
\end{equation}
where the last quantity is in the limit $n/k \ll 1$. 

Let us now calculate the same quantities in the non-commutative
description. First the energy of $n$ D0-branes not yet forming a
D2-brane can be obtained by evaluating the action on $B_a=0$ and is
given by 
\begin{equation}
n T_{(0)} {(2\pi\a)^2\over (k\a)^2} {1\over {\rm Dim}} {\rm Tr}
{1\over 6} (Y)^2.
\end{equation}
In the $n$-dimensional irreducible representation that we work with,
one can replace $(Y)^2$ with the value of the Casimir
$(n^2-1)/4$. We get
\begin{equation}\label{1}
n T_{(0)} {1\over 6} {\pi^2 (n^2-1)\over k^2}.
\end{equation}
The energy of a single D2-brane formed out of $n$ D0-branes can
be obtained by evaluating the action on $B^a=Y^a_{n-1}$, where
$Y^a_{n-1}$ are generators in the $n$-dimensional irreducible
representation. This gives zero. Thus, in the non-commutative
description the zero of energy is taken to be at the bound state of
$n$ D0-branes forming a D2-brane. The quantity (\ref{1}) is then to be
compared with the energy difference (\ref{diff1}). In the limit $n/k
\ll 1$ and $n\gg 1$ ensuring that the ordinary BI description is
acceptable, the two agree. Note that the factor of $1/6$ in the
non-commutative action receives the interpretation of $1/3!$ coming
from the $\sin$ in the ordinary BI result. 

The second quantity that must be calculated is the binding energy. It 
is to be compared to the binding energy (\ref{diff2}). Let us however 
calculate a more general quantity, namely 
the energy of the solution (\ref{sol-B}) describing $l$ D0-branes. We
have
\begin{eqnarray}
\lefteqn{
(n+l) T_{(0)} {(2\pi\a)^2\over (k\a)^2} \frac{1}{n+l} {\rm Tr}
{1\over 6} \left( (Y_{n+l-1})^2 - (Y_{n-1})^2 \right)}
\nn
\\ && =
(n+l) T_{(0)} {\pi^2\over k^2} {1\over 6} 
\left( (n+l)^2-1 - (n^2-1){n\over n+l} \right) 
\nn \\
&& \approx
{l T_{(0)} \over 2} \left( {\pi n\over k} \right)^2,
\end{eqnarray}
which is exactly the number of D0-branes $l$ times the BI result
(\ref{diff2}) for the binding energy, as one expects.
We thus showed the perfect matching between the energies computed in 
the non-commutative description and the expectations from the BI
theory. 


\section{Fluctuations and Stability}
\label{sec:stab}

As we argued in the previous section, the solutions considered have
the interpretation in terms of D0-branes. In the flat space, a
D0-brane located close enough to a (flat) D2-brane is known to be
unstable which is manifested by the presence of a tachyonic mode in
the string spectrum. The final stable configuration in this case is
the D0-brane absorbed by the D2-brane, that is, the D0-D2 bound state. 
We expect a similar phenomenon to be present in our case. One can see
whether or not a tachyonic mode is present by studying small
fluctuations around  the solution in question. Such a study was
performed in the flat case  in \cite{Strom}, and here we present a
similar analysis for our case. 

To study fluctuations, we need an expression for the second variation
of the action. A simple computation shows that it is given by
\begin{equation}
\delta^2 S = {\rm Tr} \left( 
- [\delta B_a, B_b] [\delta B_a, B_b] + ([\delta B_a, B_a])^2 +
2 i [\delta B_a, \delta B_b] F_{ab} \right).
\end{equation}
This expression generalizes the one given in \cite{Alekseev} to the
case of $F\neq 0$. 

We would like to study fluctuations around the soliton describing a
single D0-brane plus a spherical D2-brane. One can always put the
center of the D2-brane to the origin of $S^3$. The corresponding
solution then is given by
\begin{eqnarray}
  {B}_a = 
\left(
  \begin{array}{cc}
  c_a & \vec{0}\;{}^T \\
\vec{0} & {Y}_a
  \end{array}
\right)
\label{sol}
\end{eqnarray}
in the matrix notation. Here ${Y}_a$ are ${\rm SU}(2)$ generators
in the $n+1$-dimensional irreducible representation, that is, the
representation of spin $n/2$. Thus, we are describing a D2-brane made
out of $(n+1)$ D0-branes here. The constant $c_a$ 
parameterizes the moduli of the solution, and has the interpretation 
of a position of the D0-brane. Note that 
$c_a$ is a real vector due to the Hermiticity of the matrix $B^a$. The
spherical D2-brane, represented by the   
part ${Y}_a$ in the above matrix, is a spherical shell of a definite
radius, depending on $n$. We expect that the above soliton has a tachyonic 
fluctuation in a certain limited region  of moduli space of $c_a$. This
region should corresponds to the situation that the D0-brane sits very
close to the shell of the D2-brane.

Experience with the flat case teaches us that the tachyonic mode comes 
from the excitations of the string connecting the D0-brane and
D2-branes. As in the works \cite{Nikita,Nikita-Dynamics,Strom} on the
spectrum of fluctuations 
around non-commutative solitons in flat case, the 0-2 string modes are 
described by the off-diagonal elements of the matrix representing the
soliton. Hoping that the same is true in our case, let us turn on 
only the off-diagonal fluctuations. Thus, we consider the following 
perturbation of the solution
\begin{eqnarray}
\delta  {B}_a = 
\left(
  \begin{array}{cc}
  0 & v_a^\dagger \\
   v_a & 0
  \end{array}
\right)
\end{eqnarray}
where $v_a$ is a complex vectors (column) of dimension $n+1$.
Substituting this fluctuation mode into the action, we get
\begin{eqnarray}
  \delta^2 S = 
   v_a^\dagger ({Y}_b - c_b \unit) ({Y}_b - c_b \unit) v_a
- v_a^\dagger ({Y}_a-c_a\unit)({Y}_b-c_b\unit)v_b
+ i\epsilon_{abc}c_a(v_b^\dagger v_c - v_c^\dagger v_b)
\end{eqnarray}
We can use the rotation symmetry on $S^3$ to fix the value of $c_a$ as 
$c_a = \delta_{a3} c$, without losing generality. The above expression
is a Hermitian quadratic form in the complex vector space of dimension
$3(n+1)$ (3 complex vectors $v_a$). To facilitate its diagonalization,
let us define the following complex linear combinations of these 3
vectors 
\begin{equation}
\v = v_3, \qquad \v_+ = {1\over\sqrt{2}}(v_1+i v_2), \qquad
\v_- = - {1\over\sqrt{2}}(v_1-i v_2).
\end{equation}
The minus in front of the expression for $\v_-$ is for uniformity with
the similar definition of $Y_-$, see Appendix. 
Let us at the same time introduce the usual complex linear combinations
of ${\rm SU}(2)$ generators, see the Appendix for our conventions on 
normalization. We shall denote these ``raising and lowering'' operators
by $Y, Y_{\pm}$. The action, written in terms of these quantities,
becomes 
\begin{eqnarray} 
\nonumber
\frac14 \delta^2 S 
&=& \left( \dv \v + \dv_+ \v_+ + \dv_- \v_- \right)\left( 
{n\over 2}\left( {n\over 2} + 1 \right)  + c^2 \right) 
-2c\left( \dv Y \v + \dv_+ Y \v_+ + \dv_- Y \v_- \right) 
\\ \label{action-1}
&-&\left(\dv Y + \dv_+ Y_+ + \dv_- Y_-\right)
\left(Y \v - Y_+ \v_- - Y_- \v_+\right)
\\ \nonumber
&+& c \left(\dv Y + \dv_+ Y_+ + \dv_- Y_-\right)\v 
+ c \dv \left(Y \v - Y_+ \v_- - Y_- \v_+\right)
- c^2 \dv \v \\ \nonumber
&+& 2c\left(\dv_+ \v_+ - \dv_- \v_-\right).
\end{eqnarray}
Here, to write the first term, we have used the fact 
that the $(n+1)$-dimensional complex vector space 
that the vectors $\v,\v_{\pm}$ are elements of can be thought of as
the irreducible representation space. Thus, the operator $Y_a Y_a$
can be replaced by the value of the Casimir times the identity
operator. To diagonalize this quadratic form, let us introduce the
basis of eigenvectors $|m\rangle$ of the operator $Y$, and decompose
the vectors $\v, \v_{\pm}$ with 
respect to this basis. Using conjugation of generators $Y, Y_{\pm}$
by elements of ${\rm U}(n+1)$ we can always choose the highest vector
$|n/2\rangle$ to point along $\v$. The decomposition then becomes:
\begin{equation}
\v = v \, |n/2\rangle, \qquad \v_+ = \sum_m v_+^m \, |m\rangle,
\qquad \v_- = \sum_m v_-^m \, |m\rangle.
\end{equation}
Substituting these decompositions into (\ref{action-1}), and using
the standard expressions for the action of $Y_{\pm}$ on vectors
$|m\rangle$, see Appendix, one can diagonalize the quadratic form in
question rather straightforwardly. The action takes a block diagonal
form consisting of blocks not larger than $2\times 2$. The coupling of
the modes to each other is as follows. The mode $v$ gets coupled only
to $v_-^{n/2-1}$. This part of the action is 
\begin{equation}
|v|^2 \left( {n\over 2}\right) + 
\bar{v} v_-^{n/2-1} \sqrt{{n\over 2}}\left({n\over 2} - c\right) +
c.c. 
+ |v_-^{n/2-1}|^2  \left({n\over 2} - c\right)^2.
\end{equation}
This quadratic form can be easily diagonalized with the eigenvalues
being 
\begin{equation}
\lambda_0 = 0, \qquad \lambda_+^{n/2} = c^2 - nc + {n\over 2}
\left( {n\over 2}+1 \right). 
\end{equation}
The notation for the last eigenvalue will become clear when we
consider the other modes.
The modes $v_-^{n/2}$ and $v_+^{-n/2}, v_+^{-n/2+1}$ do not couple to
other modes. We get for each of these modes correspondingly
\begin{eqnarray}
&{}&|v_-^{n/2}|^2 \left( c^2 - 2c\left({n\over 2}+1\right) +
{n\over 2}\left({n\over 2}+1\right) \right), \\
&{}&|v_+^{-n/2}|^2 \left( c^2 + 2c\left({n\over 2}+1\right) +
{n\over 2}\left({n\over 2}+1\right) \right), \\
&{}&|v_+^{-n/2+1}|^2 \left( {n\over 2} + c\right)^2.
\end{eqnarray}
Thus, the corresponding eigenvalues are
\begin{eqnarray}
\lambda_{t} &=& c^2 - 2c\left({n\over 2}+1\right) +
{n\over 2}\left({n\over 2}+1\right), \\
\tilde{\lambda}_{t} &=& c^2 + 2c\left({n\over 2}+1\right) +
{n\over 2}\left({n\over 2}+1\right), \\
\lambda_-^{-n/2} &=& \left( {n\over 2} + c\right)^2.
\end{eqnarray}
We have introduced a special notation for the first two eigenvalues
because, as we shall see, they correspond to tachyonic modes.
The remainder of the action couples, in pairs, the modes $v_+^{m+1}$
and $v_-^{m-1}$, with $m=(n/2-1), (n/2-2), \ldots, (-n/2+1)$:
\begin{eqnarray}
\nonumber
\lefteqn{
|v_+^{m+1}|^2 \left( c^2 - 2cm + \Half m(m+1) 
+\Half \left( {n\over 2}\right) \left({n\over 2}+1 \right) \right)
} \\
&&+\bar{v}_+^{m+1} v_-^{m-1} 
\Half \sqrt{\left(\left({n\over 2}+1 \right)^2-m^2\right)
\left(\left({n\over 2}\right)^2 - m^2 \right)} + c.c. \\ \nonumber
&&\hs{10mm}+ |v_-^{m-1}|^2 \left( c^2 - 2cm + \Half m(m-1) 
+\Half \left( {n\over 2} \right) \left({n\over 2}+1 \right) \right).
\end{eqnarray}
The corresponding eigenvalues are
\begin{equation}
\lambda_-^m = (c-m)^2, \qquad 
\lambda_+^m = c^2 - 2mc +  {n\over 2} \left({n\over 2}+1 \right).
\end{equation}
Here we have introduced the notation $\lambda_{\pm}$ for two different
sets of eigenvalues. To summarize, the set of eigenvalues consists of: 
(i) $\lambda_{t},\tilde{\lambda}_{t}$ which correspond to tachyonic
modes; (ii) zero mode $\lambda_0$; (iii) two sets $\lambda_{\pm}^m$,
with  $m = n/2, n/2-1, \ldots, -n/2 +1$ for $\lambda_+^m$ (no
$m=-n/2$) and $m = n/2-1, \ldots, -n/2+1, -n/2$ for $\lambda_-^m$ (no
$m=n/2$). 

As we have said, the eigenvalues $\lambda_t,\tilde{\lambda}_t$
correspond to tachyonic modes. Indeed, in the region of the moduli
space 
\begin{equation}
c \in \left( {n\over 2}+1 - \sqrt{{n\over 2}+1} , \;
{n\over 2}+1 + \sqrt{{n\over 2}+1}
\right)
\end{equation} 
the eigenvalue $\lambda_t$ is negative, and in the region 
\begin{equation}
c \in \left( -\left({n\over 2}+1\right) - \sqrt{{n\over 2}+1} ,\; 
-\left({n\over 2}+1\right) + \sqrt{{n\over 2}+1}
\right)
\end{equation}
the eigenvalue $\tilde{\lambda}_t$ is negative. These values of 
$c$ correspond, as expected, to two intervals on the $z$-axes, 
near the intersection of the axes with the spherical D2-brane shell. 
Indeed, let us convert the distances measured by $c$ into physical
distances measured with respect to the closed string metric $g$.
The conversion is
\begin{equation}
({\rm distance}) = 2\pi\a {c\over \sqrt{k\a}}.
\end{equation}
The region of instability is then centered around the value
$c=\pm(n/2+1)$, which corresponds, in physical units, to the points
the distance 
\begin{equation}
R = \pi (n+2) \sqrt{{\a\over k}}
\end{equation}
away from the origin. This is the correct radius of the fuzzy
D2-brane, as predicted by the commutative description
\begin{equation}
R = \sqrt{\a k} \sin{\pi n\over k} \approx \pi n \sqrt{{\a\over k}}.
\end{equation}

As far as the instability is concerned, it appears at the 
physical distance of 
\begin{equation}\label{L}
L_{\rm instab} = (2\pi R)^{1/2} \left({\a\over k}\right)^{1/4}
\end{equation}
away from the shell. Here we have used the radius of the shell $R$ 
so as to eliminate the $n$ dependence. One would like to
compare this with the flat case result $L_{\rm instab}\sim\sqrt{\a}$. 
We see that the result (\ref{L}) is {\it different} from the flat
result in the  regime $R\ll \sqrt{k\a}$ where one can neglect the
curvature of $S^3$  and the fuzzy sphere description is valid. If, for
some reason,  the result (\ref{L}) can be extrapolated beyond its
region of validity into the region $R\sim \sqrt{k\a}$ of radii of
$S^2$ being of the  order of the radius of curvature of $S^3$, than
$L_{instab}$ equals to the flat case quantity for $R=\sqrt{k\a}$.

Note that each
tachyonic eigenvalue $\lambda_t$ or $\tilde{\lambda}_t$ represents one 
complex, or equivalently two real tachyonic modes. These have the
interpretation of coming from the two different orientations of the 
0-2 string. 

Let us now discuss the interpretation of other eigenvalues. The zero
mode appears because some symmetry is left unbroken in the solution
under consideration. Indeed, we still have a ${\rm U}(1)$ subgroup of
the global ${\rm SU}(2)$ rotating the generators. This subgroup
generates rotations about the axes of symmetry ($z$) of our
solution. There is, in addition, another ${\rm U}(1)$, which is a
subgroup of ${\rm U}(n+2)$ acting on our matrices by conjugation. This
${\rm U}(1)$ multiplies the rank $1$ and $(n+1)$ blocks of the solution
by the complex conjugate phase factors. These two ${\rm U}(1)$
correspond to two real zero modes in the fluctuation spectrum.

All other eigenvalues are positive for any $c$, as can be easily
checked. The corresponding masses thus constitute a prediction for the
low energy spectrum of 0-2 worldsheet CFT on $S^3$ in the SW limit. In
the flat case, these spectrum can be easily calculated on the CFT
side, see \cite{Strom}, and one finds  an exact agreement with the
non-commutative prediction. It would be somewhat harder to do a
similar CFT calculation in our case of $S^3$, and we shall not attempt
it here.


\section{Higgs Mechanism}
\label{sec:higgs}

Our results on stability have potentially interesting implications 
as to scenarios of gauge symmetry breaking. Let us assume that the
manifold $M_7$, which is transverse to $S^3$, contains some flat
directions, i.e., is of the type $\R^{p+1}\times
\widetilde{M}_{6-p}$. Let us have a stack of $N$ D$p$-branes, such
that the directions along the branes are in $\R^p$. One can then get
an unusual gauge symmetry breaking mechanism by separating the branes
in $S^3$. 

The low energy excitations of a stack of D$p$-branes in $\R^{10}$ are 
described by the usual action
\begin{eqnarray}
S = T_{(p)} (2\pi\alpha')^2 \int \! d^{p+1}x {\rm Tr}
\left(
\frac14 f_{\mu\nu}^2 + \frac12 (D_\mu X_a)^2 - \frac14 [X_a, X_b]^2
\right).
\end{eqnarray}
Here $f_{\mu\nu}$ is the curvature of the connection $a_\mu$,
with Greek indices corresponding to directions along the branes,
and $X_a$ are transverse scalars. 
When one puts the branes on $S^3\times\R^{p+1}\times
\widetilde{M}_{6-p}$ as described, 
the last term in this action must be replaced by the action
(\ref{redefac}) of the gauge theory on fuzzy sphere, with the
identification 
\begin{eqnarray}
  X_a = B_a/\sqrt{k \alpha'}.
\end{eqnarray}
There is also the commutator term for other transverse coordinates
when $p<6$, but we will not consider it. We thus get
\begin{eqnarray}
\label{a-1}
S = T_{(p)} (2\pi\alpha')^2 \int \! d^{p+1}x {\rm Tr}
\left(
\frac14 f_{\mu\nu}^2 + \frac12 {1\over k\a} (D_\mu B_a)^2 \right) +
\int \! d^{p+1}x \, S[B],
\end{eqnarray}
where $T_{(0)}$ in $S[B]$ must be replaced with $T_{(p)}$.

Let us first analyze the case of two D$p$-branes, when all fields are 
$2\times 2$ matrices. As we know, there are two main configurations 
which locally minimizes $S[B]$ in this case: one corresponding to branes
simply separated in $S^3$ (commuting $B^a$), and the other
corresponding to a fuzzy sphere. In the first case one can take only
$B_3$ to be non-zero, and given by $B_3 = c\sigma_3/2$. This gives the 
usual D-brane Higgs mechanism. For such $B_a$, the second term
in the action breaks symmetry down to ${\rm U}(1)$. Decomposing
$a_\mu = \sum_i a^i_\mu\sigma_i/\sqrt{2}$ we get the masses of the
off-diagonal gauge bosons $a^1_\mu,a^2_\mu$ to be 
\begin{equation}
\label{mass-1}
m = {c\over\sqrt{k\a}}.
\end{equation}
Converting $c$ into the physical distance between the D0-branes as
measured with respect to the closed string metric $L = 2\pi\a
c/\sqrt{k\a}$, we get 
\begin{equation}
m = {L\over 2\pi\a}.
\end{equation} 
This is, of course, just the energy of a fundamental string of length $L$,
as expected.

However, for small $L$ this mechanism should be modified. Indeed, as
D0-branes become too close to each other, the configuration becomes
unstable, and the fuzzy sphere forms. The value of $c$ for which this
happens can be found from the results of the previous section. For
this one must substitute $n=0$ in the formula for the tachyonic
eigenvalues. This corresponds to a system of two D0-branes. We then
get that the distance between the D0-branes (in the unphysical metric)
when the instability occurs is $2$. Thus, one must take $c=2$ in
(\ref{mass-1}). This corresponds to the physical distance between 
the branes 
\begin{equation}
L_{\rm instab} = {4\pi\a\over\sqrt{k\a}}
\end{equation}
and the mass 
\begin{equation}
\label{mass-2}
m_{\rm instab} = {2\over\sqrt{k\a}}.
\end{equation}
This is the smallest mass of the off-diagonal gauge bosons that the
$a^3_\mu$ boson is still massless. As one further decreases the
distance between the branes the fuzzy sphere forms which breaks the
gauge symmetry completely. It is easy to find the masses of gauge
bosons after the symmetry is broken. They are again obtained from the
second term in the action (\ref{a-1}), evaluated on the configuration
$B_a=Y_a=\sigma_a/2$. Using the same decomposition of the connection
into Pauli matrices as above, we get all three masses to be equal to
\begin{equation}\label{mass-3}
m = {\sqrt{2}\over\sqrt{k\a}}.
\end{equation}
This is by $\sqrt{2}$ smaller than the smallest mass $m_{\rm instab}$
that can be achieved in the usual Higgs mechanism. One can say that
some of the mass of $a^1,a^2$ went into $a^3$. 

Thus, there is the smallest mass (\ref{mass-2}) that can be achieved
by the usual Higgs mechanism. When one tries  
to further decrease it decreasing the separation between the D0-branes
the system becomes unstable and forms the fuzzy sphere, which breaks
the gauge symmetry completely. All three gauge bosons become massive
with the mass given by (\ref{mass-3}). Note interestingly, while the
local gauge symmetry is broken completely, the global symmetry is
enhanced to the full ${\rm SU}(2)$ as compared to the global symmetry
of ${\rm U}(1)$ in the usual Higgs breaking mechanism. Thus,
summarizing, if the transverse coordinates form $S^3$, there is no
usual restoration of gauge symmetry by decreasing the separation
between the branes. The only way to restore the gauge symmetry is to
take the flat space limit $\sqrt{k\a}\to\infty$. 

Other symmetry breaking patters can be obtained by considering a fuzzy
sphere of size $n$ plus a single D0-brane. When the D0-brane is far
enough from the shell, the ${\rm U}(n+1)$ symmetry of the system is
broken to ${\rm U}(1)$. As one decreases the distance between the shell
and the D0-brane, the system becomes unstable and the fuzzy sphere
of larger radius forms. The gauge symmetry is then broken completely.
Masses of gauge bosons before and after the formation of the fuzzy
sphere can be found similarly to the above calculation.


\section{Conclusions and Discussion}
\label{sec:disc}

In this paper we identified a simple subset of solutions of gauge
theory on fuzzy sphere and gave their interpretation in terms of
D0-branes located off a D2-brane. We confirmed this interpretation by
comparing the energy obtained in non-commutative and commutative
descriptions. We also looked at the spectrum of fluctuations about the
simplest solution containing a single D2-brane and D0-branes. We have
found a tachyonic mode when the D0-brane is located close enough to
the shell of the D2-brane. This is similar to the flat case, although
the expression for the distance when the instability occurs depends on
the radius of the  D2-brane shell and the radius of $S^3$ and is thus
different from the  flat case result. We have also discussed how this
instability modifies the usual Higgs mechanism for small separation
between the branes.

Since the instability occurs when D0-brane is too close to the shell
of D2-brane, it is natural to expect that a similar instability is
present in a system of several non-cocentric D2-branes, in the case
when the shells are too close to each other. However, this case would
be much harder to analyze quantitatively, and we did not consider it
in this paper. 

Considering the fluctuations we have analyzed only the 0-2 part of the 
spectrum, which is the most interesting because it is this part that
contains the tachyonic modes. It would be interesting to compare our
prediction for the 0-2 spectrum with a direct CFT calculation for
0-2 string in $S^3$ in the SW limit. Another possible calculation is
to consider  
0-0 and 2-2 modes. For the later case there is a prediction from
the usual BI theory in \cite{Bachas}. These fluctuations in the 
non-commutative description were recently studied in \cite{Japan},
but no diagonalization was given. Thus, it is still an open
problem to match the non-commutative spectrum of 2-2 fluctuations
to the BI prediction of \cite{Bachas}. 

As is discussed in \cite{Alekseev}, and more recently in \cite{Japan}, 
the non-commutative gauge theory action we considered is a bosonic
part of certain supersymmetric theory on the fuzzy sphere. Although
the SUSY transformation law for fermions was not given in
\cite{Alekseev}, it can be guessed by using the fact that the action
is the sum of YM and CS terms. Indeed, for the case of commutative
(but not necessarily Abelian) YM CS gauge theory, the supersymmetry
transformation for the gaugino is \cite{Lee,French} 
\begin{eqnarray}
  \delta \lambda
= {1\over 2} F_{ab} \gamma^{ab} \eta
\label{susy}
\end{eqnarray}
where $\eta$ is an infinitesimal Dirac spinor parameter, $F_{ab}$ are
the 
field strength, and $\gamma^{ab}$ are the usual commutators of 3D 
$\gamma$-matrices. A natural way to generalize it to the
non-commutative case is to replace the field strength $F$ by its
non-commutative version (\ref{defF}). This would be consistent with
the claim of  \cite{Alekseev-Old-1,Alekseev-Old-2,Bachas} that
spherical D2-branes constitute supersymmetric configurations, for the
field strength for them vanishes, and the gaugino is SUSY
invariant\footnote{ 
Although the translation of the spherical D2-brane in $S^3$ shifts
$F$ to a non-zero value and thus seems to
break the supersymmetry, this could be restored by
introducing non-linearly realized supersymmetries that are usually
present in the D-brane actions \cite{nonlinear, hashimoto}.
}. However, the 
transformation law (\ref{susy}) with $F$ given by (\ref{defF}) is
not what is claimed in \cite{Japan} to be the correct law for SUSY
gauge theory on the fuzzy sphere. There, the authors had proposed a
SUSY extension of the bosonic action with the transformation law using
the usual $F_{ab} = \partial_a A_b -\partial_b A_a + [A_a,A_b]$. This
field strength does not vanish for a fuzzy sphere, which makes it not
a SUSY configuration. A possible resolution of this apparent
contradiction might be that there are several different SUSY
extensions of the bosonic action (\ref{redefac}), and in the context
of D-branes the one with the transformation law (\ref{susy}) with $F$
given by (\ref{defF}) should be used. Such a SUSY extension is not the
one given in \cite{Japan}, and it remains to be seen whether it
exists. 


\appendix
\section{${\rm SU}(2)$ operators}

We use the following definition of ``raising and lowering''
${\rm SU}(2)$ operators:
\begin{equation}
Y = Y_3, \qquad Y_+ = {1\over\sqrt{2}}(Y_1+i Y_2), \qquad
Y_- = - {1\over\sqrt{2}}(Y_1-i Y_2).
\end{equation}
Our conventions are the same as those of \cite{VK}. The action of 
$Y_\pm$ on eigenvectors of $Y$, in the $(n+1)$-dimensional
irreducible representation, is
\begin{eqnarray} 
Y_+ |m\rangle &=& \sqrt{{1\over2}\left({n\over 2}-m\right)
\left({n\over 2}+m+1\right)} |m+1\rangle, \\
Y_- |m\rangle &=& - \sqrt{{1\over2}\left({n\over 2}+m\right)
\left({n\over 2}-m+1\right)} |m-1\rangle.
\end{eqnarray}


\vs{10mm}
\noindent
{\large \bf Acknowledgments}

K. K. would like to thank J.\ David and G.\ Horowitz for discussions.
K. H. was supported in part 
by Japan Society for the Promotion of
Science under the Postdoctoral Research Program (\# 02482).
K. K. was supported by the NSF grant PHY95-07065.

\newcommand{\J}[4]{{\sl #1} {\bf #2} (#3) #4}
\newcommand{\andJ}[3]{{\bf #1} (#2) #3}
\newcommand{\AP}{Ann.\ Phys.\ (N.Y.)}
\newcommand{\MPL}{Mod.\ Phys.\ Lett.}
\newcommand{\NP}{Nucl.\ Phys.}
\newcommand{\PL}{Phys.\ Lett.}
\newcommand{\PR}{ Phys.\ Rev.}
\newcommand{\PRL}{Phys.\ Rev.\ Lett.}
\newcommand{\PTP}{Prog.\ Theor.\ Phys.}

\end{document}